\documentclass[prl,twocolumn,amsmath]{revtex4}
\usepackage{graphicx}
\usepackage{bm}
\usepackage{subfigure}
\usepackage{amsmath}
\usepackage[utf8]{inputenc}

\newcommand{\pr}{Phys. Rev.\ }

\newcommand{\jpa}{J. Phys. A\ }
\newcommand{\jpb}{J. Phys. B\ }

\newcommand{\etal}{{\em et al. }}
\newcommand{\etals}{{\em et al.}}

\begin{document}

\title{Comment on ``Unity-Efficiency Parametric Down-Conversion via Amplitude Amplification''}

\author{M.~K. Olsen}
\affiliation{School of Mathematics and Physics, University of Queensland, Brisbane, 
Queensland 4072, Australia.}

\date{\today}

\maketitle

In a recent letter, Niu \etals~\cite{Shapiro} have proposed a method for increasing the efficiency of spontaneous parametric down-conversion (SPDC), using input Fock states and amplitude amplification. Their theoretical treatment uses a method pioneered by Walls and Tindle~\cite{ChrisDan} for solving the state coefficients in the Schödinger equation for second harmonic generation, and also used by Podovshedov \etals~\cite{Podoshvedov} to find analytical solutions for small number input Fock states.

The authors say ``We construct the SPDC solution for an arbitrary single-mode pure state pump as an iteration that we can evaluate numerically for pump photon numbers up to 50.'' This was surprising because the same method can solve the Bose-Hubbard dimer model for hundreds of atoms without any problems. In fact, I was able to solve the SPDC dynamics for up to 10$^4$ input photons in a Fock state on a medium performance laptop. The positive-P representation~\cite{P+} can be used to solve the same system for an almost arbitrary number of input photons, with Fock states able to be represented exactly~\cite{WPstates}. The truncated Wigner representation~\cite{Graham} can also solve for the dynamics of this system with input Fock states, with an error that scales as $1/n^{2}$, where $n$ is the number of photons in the Fock state.

\begin{figure}[tbhp]
\includegraphics[width=0.75\columnwidth]{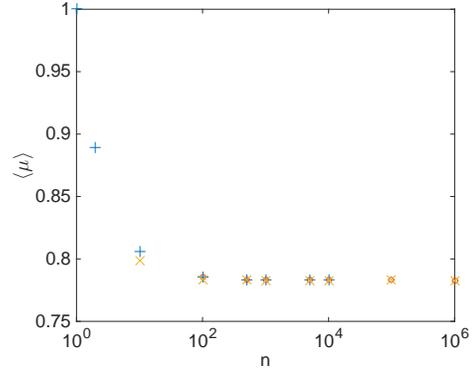}
\caption{(colour online) Expectation value of the efficiency from the matrix method (+), the positive-P (o), and the truncated Wigner approximation (x).}
\label{fig:muvek}
\end{figure}

The authors introduce a definition of  ``total quantum conversion efficiency'' in their Eq. 7, which is actually an averaged conversion efficiency for a particular interaction time. This leads the authors to state that only the case with an input of a one photon Fock state can give complete conversion efficiency. In fact, their definition of conversion efficiency is based on the expectation number of downconverted photons, which cannot be strictly converted to an efficiency in this system. I show results for their definition, with notation $\langle\mu\rangle$, in Fig.~\ref{fig:muvek}, for up to $n=10^{6}$ in a Fock state.

My analysis is different and involves realising that means have a distribution about their expectation values. Following this approach, we see that unity efficiency is possible for almost any input number of photons for a given crystal length, although with decreasing probability as the intensity increases. In this case, the probability of unity efficiency at any given time is the maximum of
$|f_{n}^{(n)}(t)|^{2}$, in the notation used by Niu \etal This is the probability that all input photons have down-converted at time $t$, and also importantly, can only be one if no low frequency photons have upconverted before this time. We will denote the maximum value of this as $\mu_{max}$, and this can be calculated using the same method as in Niu \etal  
 
\begin{center}
 \begin{tabular}{||c || c  |  c | c | c | c | c | c | c | c ||} 
 \hline
 $n$ & 1 & 2 & 10 & 100 & 500 & 10$^{3}$ & 10$^{4}$ \\ 
 \hline
 $\mu_{max}$ & 1 & 0.8889 & 0.5243 & 0.1822 & 0.0823 & 0.0582 & 0.0184 \\ 
 \hline
\end{tabular}
\end{center}

The values are shown in the table above, from which can be seen that with $n=2$, unit efficiency can be obtained $89\%$ of the time. Even with $n=10^{4}$, unit efficiency is possible with a non-zero probability. My view is that examining the full number statistics of the outputs means that the method proposed by Nui \etal will not be as effective as claimed.

\end{document}